\documentclass[prb,aps,showpacs,floats,twocolumn]{revtex4}
\usepackage{graphicx}
\usepackage{amsmath}

\newcommand{\dagga}{{\phantom{\dagger}}}

\begin{document}
\title{Superconductivity from spoiling magnetism in the Kondo lattice model}
\author{Mohammad Zhian Asadzadeh, Michele Fabrizio, and Federico Becca}
\affiliation{Democritos Simulation Center CNR-IOM Istituto Officina dei Materiali and International
School for Advanced Studies (SISSA), Via Bonomea 265, 34136 Trieste, Italy}

\begin{abstract}
We find evidence that superconductivity intrudes into the paramagnetic-to-magnetic transition of the Kondo 
lattice model if magnetic frustration is added. Specifically, we study by the variational method the model on 
a square lattice in the presence of both nearest- ($t$) and next-nearest-neighbor ($t^\prime$) hopping of 
the conduction electrons. We find that, when $t^\prime/t>0$, a $d$-wave superconducting dome emerges 
between the magnetic and paramagnetic metal phases and close to the compensated regime, i.e., the number of 
conduction electrons equals the number of localized spin-1/2 moments. Superconductivity is further 
strengthened by a direct antiferromagnetic exchange $J_H$ between the localized moments, to such an extent 
that we observe coexistence with magnetic order.
\end{abstract}


\maketitle

\section{Introduction}\label{sec:intro}

The emergence of superconductivity in strongly-correlated electron systems has become quite a common 
phenomenon, observed by now in a wealth of different materials that include also so-called heavy-fermion 
compounds, where unconventional superconductivity was first reported in CeCu$_2$Si$_2$.~\cite{steglich1979}
The characteristic properties of heavy fermions derive from the coexistence of itinerant electrons and 
localized moments residing on partly filled $f$-shells of rare earth or actinide ions. The pairing mechanism 
in heavy fermions has been the subject of an intense debate, also in connection with high-temperature 
superconductors,~\cite{norman2011,scalapino2012} with which heavy fermions share unconventional pairing 
symmetry. For instance, recent scanning tunnelling spectroscopy on CeColn$_5$ unveiled the presence 
of nodal points in the superconducting gap compatible with a $d_{x^2-y^2}$ symmetry,~\cite{allan2013} just 
like in cuprates. The widely accepted picture is that pairing is mediated by spin fluctuations of a nearby 
antiferromagnetic phase, as suggested for CePd$_2$Si$_2$ and CeIn$_3$.~\cite{mathur1998,grosche2001} 
This view is supported by the evidence that superconductivity in heavy fermions almost always appears in the 
vicinity of the quantum critical point that separates a paramagnetic metal phase from a magnetically-ordered 
one.~\cite{si2010} Even more remarkably, some compounds show a coexistence of magnetism and superconductivity, 
as observed in CeRhSi$_2$,~\cite{movshovich1996} CeRhIn$_5$,~\cite{yashima2007} and 
CeCo(In$_{1-x}$Cd$_x$)$_5$.~\cite{nair2010} 

From the theoretical side, the Kondo lattice model (KLM), which has been introduced by Doniach in
1977,~\cite{doniach1977} is believed to capture the basic properties of heavy fermions. While in one spatial 
dimension its physical behavior is well understood,~\cite{tsunetsugu1997} the more relevant two- and 
three-dimensional cases are much less known, especially concerning (possible) superconducting properties. 
Most of the analytical understanding is based upon slave bosons and large-$N$ approaches,~\cite{coleman2007} 
which find that $d$-wave superconductivity can be indeed stabilized in two dimensions through the 
resonating-valence bond (RVB) mechanism, similarly to what has been suggested long ago by Anderson for 
high-temperature superconductors.~\cite{anderson1987} As far as numerical calculations are concerned, 
unfortunately quantum Monte Carlo methods suffer from sign problems away from the compensated regime (where 
the number of itinerant electrons equals the number of localized spin-1/2 moments), while exact diagonalization 
and density-matrix renormalization group (DMRG) are limited to small clusters. Nonetheless, DMRG calculations
suggest that the standard KLM does not support superconductivity.~\cite{xavier2003} This is also the conclusion 
of variational Monte Carlo calculations, which show that $d$-wave pairing is indeed present in the paramagnetic 
sector of the KLM, but it is easily defeated by magnetism.~\cite{asadzadeh2013} Instead, recent dynamical 
mean-field theory (DMFT) results obtained some evidence for an unexpected $s$-wave superconductivity close to 
the compensated regime and relatively large Kondo coupling.~\cite{bodensiek2013}

In general, the weakness of superconductivity seems the consequence of the strength of magnetism, reinforced 
in the model calculation by the bipartite square lattice and by the unfrustrated hopping. Since lack of magnetic 
frustration is rather exceptional in real materials, it is worth and legitimate to investigate how frustration 
modifies the phase diagram of the KLM.~\cite{coleman2010} Frustration in real heavy-fermion materials may take 
various forms. In certain cases, it can appear as a direct geometric frustration, as in the pyrochlore material 
Pr$_2$Ir$_2$O$_7$~\cite{nakatsuji2006} and in the Shastry-Sutherland lattice compound 
Yb$_2$Pt$_2$Pb,~\cite{kim2008} in others by competing interactions of various kinds. 
The role of frustration in the KLM has been investigated in different works, especially focusing on magnetic
properties.~\cite{motome2010,bernhard2011,peters2011,rau2014} More recently, dynamical cluster approximation 
(DCA) calculations  on the periodic Anderson model also suggested that frustration may lead to $d$-wave 
superconductivity in the vicinity of an antiferromagnetic quantum critical point.~\cite{wu2014}

There is another ingredient worth to be included to better reproduce the phase diagram of heavy fermions.  
Realistically, one expects that $f$-electrons are mutually coupled mostly through the conduction electrons, 
i.e., via the Ruderman-Kittel-Kasuya-Yosida (RKKY) exchange. However, most of the approximate methods used to 
study the KLM are unable to account for the RKKY interaction, unless long-range magnetic order is explicitly 
assumed. This inserts a bias in the calculations, since magnetic solutions can profit from the RKKY interaction,
while non-magnetic ones cannot even exploit short-range magnetic correlations, e.g., to stabilize 
superconductivity. A way to mitigate this flaw is to add a direct $f{-}f$ exchange $J_H$ mimicking the actual 
RKKY interaction. The role of $J_H$ for reproducing the magnetic properties of heavy fermions has been 
highlighted several times.~\cite{coleman1989,iglesias1997,kim2003,xavier2008,liu2012,isaev2013} Moreover, 
there are also suggestions that $J_H$ may be important to understand superconductivity in 
UPd$_2$Al$_3$.~\cite{sato2001} 

In this paper, we show that the underlying superconducting properties of the Kondo lattice model naturally
emerge whenever magnetism is frustrated and/or local Cooper pairs are reinforced by a direct $f{-}f$ exchange.
Here, the mechanism leading to superconductivity can be captured by the standard approach where spin fluctuations 
mediate pairing; however, without these extra ingredients, pairing is not strong enough to overcome magnetic 
long-range order. One of the main signatures of this mechanism is the $d_{x^2-y^2}$ symmetry of the order 
parameter, which arises from the indirect coupling of conducting electrons through localized moments.

The paper is organized as follow: in Sec.~\ref{sec:model}, we introduce the KLM and describe the numerical
method, in Sec.~\ref{sec:results}, we present the results, and finally, in Sec.~\ref{sec:conc}, we draw our
conclusions.
 
\section{Model and method}\label{sec:model}

We explore the phase diagram of an extended KLM on a square lattice described by the Hamiltonian:
\begin{align}
\cal{H} =& -t\sum_{\langle i,j\rangle,\sigma} \big(c_{i,\sigma}^\dagger c_{j,\sigma} + h.c.\big)
         -t^\prime\sum_{\langle\langle i,j\rangle\rangle,\sigma} \big(c_{i,\sigma}^\dagger c_{j,\sigma}
         + h.c.\big)\nonumber\\
         & +J\sum_i {\bf S}_i\cdot {\bf s}_i +J_H\sum_{\langle i,j\rangle} {\bf S}_i \cdot {\bf S}_j,
         \label{eq:Fkondo-hamilt}
\end{align}         
where $\langle i,j\rangle$ and $\langle\langle i,j\rangle\rangle$ imply that $i$ and $j$ are nearest neighbors 
and next-nearest neighbors, respectively; $\mathbf{S}_i$ is the spin 1/2-operator of the local moment 
at site $i$, and $\mathbf{s}_i$ that of the conduction electrons. Hereafter, we shall refer to  the frustrated 
Kondo lattice model (FKLM) when $t^\prime\not =0$ but $J_H=0$, and to the Kondo-Heisenberg lattice model 
(KHLM) in the opposite case of $t^\prime=0$ but $J_H\not =0$. We take $t=1$ as the energy unit, and study the 
phase diagram in the the uncompensated regime, where the density of $c$-electrons $n_c<1$, by varying the 
frustrating hopping $t^\prime$, the Kondo coupling $J$, and the super-exchange $J_H$. 

We study the ground state of Eq.~(\ref{eq:Fkondo-hamilt}) by variational Monte Carlo technique. The variational 
wave function is defined by:
\begin{equation}
|\Psi_v \rangle= {\cal P}_f |\Phi_{\rm MF}\rangle,
\end{equation}
where ${\cal P}_f$ is the Gutzwiller projector that enforces single occupancy of $f$ electrons on each site, 
while $|\Phi_{\rm MF}\rangle$ is an uncorrelated wave function defined as the ground state of a non-interacting 
variational Hamiltonian, ${\cal H}_{\rm MF}$, that in general contains as variational parameters $c{-}c$, 
$c{-}f$ and $f{-}f$ hybridization terms, an energy shift of the $f$-orbitals, staggered magnetic fields acting 
on $c$- and $f$-electrons, as well as BCS coupling terms.~\cite{asadzadeh2013} Depending on ${\cal H}_{\rm MF}$ 
we can describe different uncorrelated states: 

1) A paramagnetic normal metal, which we denote by PM:
\begin{equation}
{\cal H}_{\rm PM}=\sum\limits_{k,\sigma}
\left[
\begin{array}{cc}
c^\dag_{k,\sigma} & f^\dag_{k,\sigma}
\end{array}
\right]
\begin{bmatrix}
 \chi^{cc}_{k} & V      \\
 V             & \chi^{ff}_{k}
\end{bmatrix}
\left[
\begin{array}{c}
c^\dagga_{k,\sigma} \\
f^\dagga_{k,\sigma}
\end{array}
\right].
\end{equation}

2) A singlet superconductor with inversion symmetry, denoted by PM+BCS:
\begin{equation}
\begin{split}
{\cal H}_{\rm PM+BCS}=\sum\limits_{k}
\left[
\begin{array}{cccc}
c^\dag_{k,\uparrow} & c^\dagga_{-k,\downarrow} & f^\dag_{k,\uparrow} & f^\dagga_{-k,\downarrow}
\end{array}
\right]
\times \\
\begin{bmatrix}
 \chi^{cc}_{k}   & \Delta^{cc}_{k} & V               & \Delta^{cf}_{k} \\
 \Delta^{cc}_{k} & -\chi^{cc}_{k}  & \Delta^{cf}_{k} & -V              \\
 V               & \Delta^{cf}_{k} & \chi^{ff}_{k}   & \Delta^{ff}_{k} \\
 \Delta^{cf}_{k} & -V              & \Delta^{ff}_{k} & -\chi^{ff}_{k} 
\end{bmatrix}
\left[
\begin{array}{c}
c^\dagga_{k,\uparrow}  \\
c^\dag_{-k,\downarrow} \\
f^\dagga_{k,\uparrow}  \\
f^\dag_{-k,\downarrow}
\end{array}
\right].
\end{split}
\end{equation}

3) An antiferromagnetic metal:
\begin{equation}
\begin{split}
{\cal H}_{\rm AF} =\sum\limits_{k\in MBZ,\sigma}
\left[
\begin{array}{cccc}
c^\dag_{k,\sigma} & c^\dag_{k+Q,\sigma} & f^\dag_{k,\sigma} & f^\dag_{k+Q,\sigma}
\end{array}
\right]
\times \\
\begin{bmatrix}
 \chi^{cc}_{k} & m_f \sigma       & V             & 0             \\
 m_f \sigma    & -\chi^{cc}_{k}   & 0             & V             \\
 V             & 0                & \chi^{ff}_{k} & m_c \sigma    \\
 0             & V                & m_c \sigma    & -\chi^{ff}_{k}
\end{bmatrix}
\left[
\begin{array}{c}
c^\dagga_{k,\sigma}   \\
c^\dagga_{k+Q,\sigma} \\
f^\dagga_{k,\sigma}   \\
f^\dagga_{k+Q,\sigma}
\end{array}
\right],
\end{split}
\end{equation}
where the sum over $k$ is restricted to the reduced Brillouin zone and $Q=(\pi,\pi)$. In the latter case, two possible
states can be variationally obtained that differ by the topology of their Fermi surface,~\cite{ogata2007} either 
electron- or hole-like, which we refer to as AFe and AFh, respectively. 

In all the previous cases, $\chi^{cc}_{k}=-2 (\cos k_x + \cos k_y) -4 \tilde{t}^\prime \cos k_x \cos k_y$ and 
$\chi^{ff}_{k}=-2 \tilde{t}_f (\cos k_x + \cos k_y) -\mu_f$; $\tilde{t}^\prime$, $\tilde{t}_f$, and $\mu_f$ being
variational parameters, as well as hybridization $V$ and magnetizations for $c$ and $f$ electrons ($m_c$ and $m_f$). 
The best variational energies are obtained allowing a superconducting pairing with 
$d_{x^2-y^2}$ symmetry, compared to on-site or extended $s$-wave symmetry. The most relevant ones are
$\Delta^{cc}_{k} = 2 \Delta_c (\cos k_x - \cos k_y)$ and $\Delta^{ff}_{k} = 2 \Delta_f (\cos k_x - \cos k_y)$
($\Delta_c$ and $\Delta_f$ being additional variational parameters).

4) Whenever antiferromagnetism coexist with superconductivity, we label the states as AFh+BCS or AFe+BCS (we do not
write the explicit form of the $8 \times 8$ matrix representing the uncorrelated Hamiltonian). 

We would like to mention that we do not consider magnetic states with incommensurate or inhomogeneous patterns, since,
in general, incommensurate spirals are fragile and do not give rise to sizable energy gains, while inhomogeneous 
phases are not expected in the antiferromagnetic Kondo lattice model (on the contrary, the latter ones have been
found in models with ferromagnetic Kondo super-exchange).~\cite{aliaga2001,garcia2004} The variational parameters 
are obtained by minimizing the total energy by quantum Monte Carlo simulations.~\cite{sorella2005} Calculations 
have been performed on $L \times L$ square clusters with $L$ ranging from $8$ to $20$.

\begin{figure}
\vspace{0.2cm}
\includegraphics[scale=0.32]{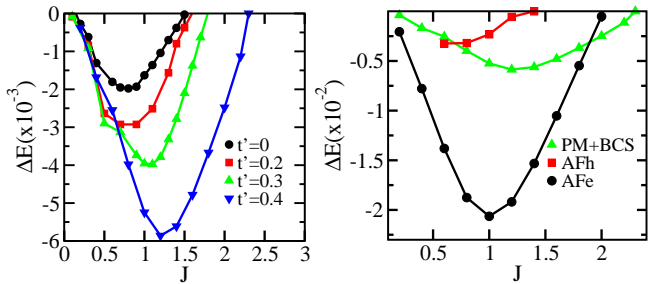}
\caption{\label{fig:condens-energy}
(Color on-line) Left panel: energy (per site) difference between the superconducting state and the paramagnetic 
metal as a function of the Kondo coupling $J$ for different values of positive $t^\prime$. Right panel: energy 
(per site) difference of magnetic and superconducting phases with respect to paramagnetic metal versus $J$ for 
$t^\prime=0.4$. The cluster has $L=8$ and the density of conduction electrons is $n_c \simeq 0.94$.}
\end{figure}

\begin{figure}
\vspace{0.4cm}
\includegraphics[scale=0.32]{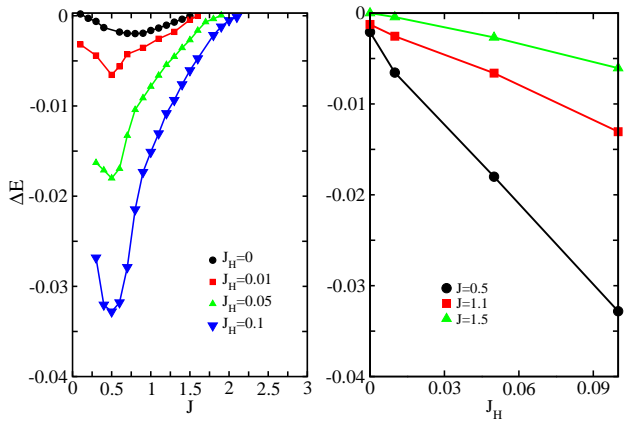}
\caption{\label{fig:pm-kh}
(Color on-line) Left panel: superconducting condensation energy (per site) versus the Kondo exchange $J$ for 
different values of $J_H$ and $n_c \simeq 0.94$. Right panel: the same quantity for fixed value of $J$ versus 
$J_H$.}
\end{figure}

\section{Results}\label{sec:results}

We start by discussing the results for the FKLM of Eq.~(\ref{eq:Fkondo-hamilt}) with $J_H=0$. First, we focus 
on the paramagnetic sector, namely on PM and PM+BCS states. We report in Fig.~\ref{fig:condens-energy} the 
condensation energy, computed as the energy difference between the optimized superconducting state and the best 
paramagnetic metal, for $n_c \simeq 0.94$ and different values of $t^\prime$. The presence of a positive 
next-nearest-neighbor $t^\prime$ is found to considerably enhance the condensation energy, while negative values 
considerably suppresses it (not shown). The enlargement of the stability region of superconductivity for 
$t^\prime>0$ is also remarkable: while for $t^\prime=0$ the condensation energy vanishes for $J \simeq 1.5$, 
for $t^\prime=0.4$ superconductivity survives up to $J \simeq 2.3$. 

Let us now consider magnetic states. We recall that, in the absence of frustration, the magnetic solution 
has always lower energy than the superconducting one, when the latter is stable, hence the actual phase diagram 
does not include superconductivity at all.~\cite{asadzadeh2013,ogata2007} This situation changes when frustration
is added. In Fig.~\ref{fig:condens-energy}, we report the optimized energy of PM+BCS, AFh, and AFe states relative
to the PM one, for $t^\prime=0.4$ and $n_c \simeq 0.94$ (in this case, we do not find any appreciable gain by 
allowing both magnetism and superconductivity). The first observation is that the AFh state is strongly hindered 
by $t^\prime$, while the AFe state lowers its energy. The most important feature is that now the superconducting 
state takes over antiferromagnetism in a wide range of parameters.

\begin{table}
\caption{\label{tab1}
Energies per site of the paramagnetic metal and the best superconducting state for $t^\prime=0.4$, $J=2.1$ and
$n_c \simeq 0.94$ on $L \times L$ clusters.}
\begin{tabular}{ccccc}
\hline
$L$ & $n_c$ & $E^{PM}$     & $E_{PM+BCS}$ & $\Delta E= E_{PM+BCS}-E_{PM}$ \\
\hline \hline
8   & 0.937 & -2.18660(2)  & -2.18912(2)  & -0.00252(4) \\
12  & 0.930 & -2.18274(2)  & -2.18429(2)  & -0.00155(4) \\
14  & 0.938 & -2.18721(2)  & -2.18866(2)  & -0.00145(4) \\
16  & 0.937 & -2.18650(2)  & -2.18784(2)  & -0.00134(4) \\
18  & 0.926 & -2.18016(2)  & -2.18126(2)  & -0.00110(4) \\
20  & 0.940 & -2.18782(2)  & -2.18910(2)  & -0.00128(4) \\
\hline \hline
\end{tabular}
\end{table}

The existence of a superconducting phase close to the paramagnetic to magnetic transition is confirmed by 
performing the size scaling up to $L=20$. In Table~\ref{tab1}, we report the energies of the paramagnetic 
metal and the best superconducting state (as well as their difference) for $t^\prime=0.4$, $J=2.1$, and 
$n_c \simeq 0.94$. Due to finite-size effects, it is not possible to fix exactly the same value of $n_c$ for all 
clusters, nevertheless, the differences in the electron concentration is very similar in all cases. Although the 
condensation energy is quite reduced from $L=8$ to $L=12$, it remains essentially constant from $L=12$ to $L=20$, 
indicating a finite value in the thermodynamic limit. In light of these results, we believe that a true 
superconducting phase does exist in the vicinity of the magnetic transition (especially for large values of $J$ 
and $t^\prime$). Finally, further away from compensated regime (i.e., for $n_c \lesssim 0.8$) the superconducting 
phase is defeated by the AFe state, and eventually disappears.

\begin{figure}
\vspace{0.2cm}
\includegraphics[scale=0.3]{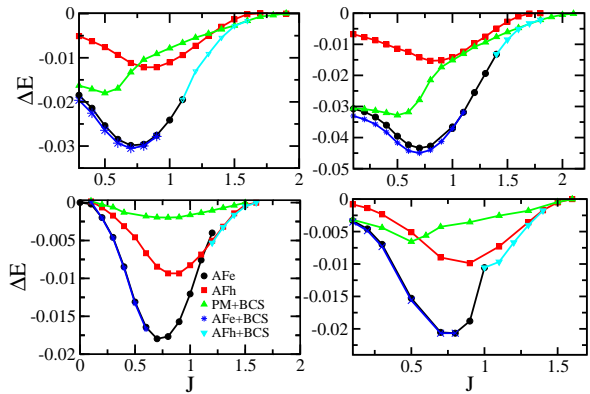}
\caption{\label{fig:all-kh}
(Color on-line) Energy (per site) difference of magnetic and superconducting phases with respect to paramagnetic 
state as a function of the Kondo exchange $J$, for different values of $J_H$ and $n_c \simeq 0.94$. The cases 
with $J_H=0$ (bottom left), $J_H=0.01$ (bottom right), $J_H=0.05$ (top left), and $J_H=0.1$ (top right) are 
reported.}
\end{figure}

\begin{figure}
\begin{center}
\includegraphics[scale=0.23]{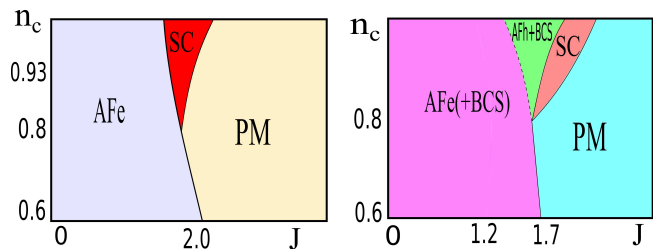}
\caption{\label{fig:phase-diagram}
(Color on-line) Schematic phase diagrams in the plane of $n_c$ and $J$ for frustrated Kondo lattice model 
(left panel, with $t^\prime=0.4$) and Kondo-Heisenberg lattice model (right panel, with $J_H=0.1$).}
\end{center}
\end{figure}

We now turn to the KHLM, with $t^\prime=0$ but $J_H>0$ in Eq.~(\ref{eq:Fkondo-hamilt}). We observe that a 
variational wave function could in principle account for the RKKY exchange, hence not require any $J_H$, through 
spin-spin Jastrow factors. However, in practice this is unfeasible unless spin $SU(2)$ symmetry is explicitly 
broken. Therefore, even though our variational approach is more accurate than Hartree-Fock, we still need to 
include a direct $f{-}f$ exchange to add magnetic short-range correlations provided in reality by the RKKY 
interaction.  

As before, we start by the paramagnetic sector. Also in this case the superconducting pairing has $d_{x^2-y^2}$ 
symmetry. In Fig.~\ref{fig:pm-kh}, we show the condensation energy for different values of $J_H$ at 
$n_c \simeq 0.94$. The case of $J_H=0$ has been also reported for comparison. The maximum gain remains peaked 
around $J=0.5$ but increases monotonically with $J_H$. Remarkably, even tiny values of $J_H$ substantially enhance
the condensation energy. The inclusion of $J_H$ not only enlarges the condensation energy but also the stability 
region of superconductivity. While at $J_H=0$ the transition to a normal metal occurs at $J \simeq 1.5$, for 
$J_H=0.1$ the superconducting state remains stable up to $J \simeq 2.2$.

To assess whether superconductivity does exist in the phase diagram, we now examine also magnetic states.
Obviously, a direct antiferromagnetic interaction $J_H$ enhances the tendency towards N\'eel order, hence 
enlarges the stability region of antiferromagnetism. In Fig.~\ref{fig:all-kh}, we show the energy of magnetic
and superconducting states relative to the paramagnetic state, for different values of $J_H$ and $n_c \simeq 0.94$. 
The case $J_H=0$ has been also included for comparison. Interestingly, upon increasing $J_H$ the superconducting 
phase finally gets energetically more favorable than AFh. The stability of a pure superconducting phase close
to the paramagnetic to magnetic transition is confirmed by a size scaling of the condensation energy, see
Table~\ref{tab2}. Also in this case, all evidences point to a condensation energy that remains finite in the 
thermodynamic limit. Therefore, a superconducting region in the vicinity of the magnetic quantum critical point 
emerges as before, this time thanks to a finite $J_H$.

\begin{table}
\caption{\label{tab2}
Energies per site of the paramagnetic metal and the best superconducting state for $J_H=0.1$, $J=1.8$ and
$n_c \simeq 0.94$ on $L \times L$ clusters.}
\begin{tabular}{ccccc}
\hline
$L$ & $n_c$ & $E^{PM}$     & $E_{PM+BCS}$ & $\Delta E= E_{PM+BCS}-E_{PM}$ \\
\hline \hline
8   & 0.937 & -1.99355(2)  & -1.99570(2)  & -0.00215(4) \\
12  & 0.930 & -1.98825(2)  & -1.98961(2)  & -0.00136(4) \\
14  & 0.938 & -1.99448(2)  & -1.99584(2)  & -0.00136(4) \\
16  & 0.937 & -1.99353(2)  & -1.99469(2)  & -0.00116(4) \\
18  & 0.926 & -1.98471(2)  & -1.98558(2)  & -0.00087(4) \\
20  & 0.940 & -1.99539(2)  & -1.99645(2)  & -0.00106(4) \\
\hline \hline
\end{tabular}
\end{table}

Furthermore, $J_H$ also stabilizes coexistence of pairing and magnetism, especially close to the quantum critical 
point. Indeed, we find a substantial energy gain when adding superconducting parameters on top of the AFh state, 
giving rise to a AFh+BCS phase. In other words, upon reducing the Kondo coupling, the paramagnetic metal first 
becomes superconducting through a second-order transition and then acquires magnetic order, still displaying a 
sizable electron pairing, see Fig.~\ref{fig:all-kh}. By further reducing the Kondo exchange, a first-order 
transition to a AFe state occurs. Its energy can be slightly lowered by allowing for a BCS coupling, see 
Fig.~\ref{fig:all-kh}. This gain decreases with increasing $L$, possibly indicating that the coexistence of 
magnetism and superconductivity will disappear in the thermodynamic limit, see Table~\ref{tab3}.

For lower electron densities, namely for $n_c \lesssim 0.8$ the AFh state cannot be stabilized anymore, similarly 
to the $J_H=0$ case. In addition, the pure superconducting phase PM+BCS is now defeated by the AFe state,
although the latter may still allow for a coexisting superconductivity.

\begin{table}
\caption{\label{tab3}
Energies per site of the magnetic state (AFe) and the one with both magnetism and superconductivity for
$J_H=0.1$, $J=0.5$ and $n_c \simeq 0.94$ on $L \times L$ clusters.}
\begin{tabular}{ccccc}
\hline
$L$ & $n_c$ & $E^{AFe}$   & $E_{AFe+BCS}$ & $\Delta E= E_{AFe+BCS}-E_{AFe}$ \\
\hline \hline
8   & 0.937 & -1.70758(2) & -1.70972(2)   & -0.00214(4) \\
12  & 0.930 & -1.69280(2) & -1.69442(2)   & -0.00162(4) \\
14  & 0.938 & -1.69733(2) & -1.69877(2)   & -0.00144(4) \\
16  & 0.937 & -1.69240(2) & -1.69353(2)   & -0.00113(4) \\
18  & 0.926 & -1.69588(2) & -1.69686(2)   & -0.00098(4) \\
20  & 0.940 & -1.69372(2) & -1.69419(2)   & -0.00047(4) \\
\hline \hline
\end{tabular}
\end{table}

\section{Conclusions}\label{sec:conc}

In summary, we have shown that magnetic frustration in the Kondo lattice model has the important role of suppressing
magnetic order hence uncovering superconductivity, which we find intrudes between the paramagnetic and 
antiferromagnetic metal phases, see Fig.~\ref{fig:phase-diagram}. Superconductivity is further stabilized by 
short-range magnetic correlations, which in reality are yielded by the RKKY exchange but which we had to enforce 
in our variational calculation through a direct antiferromagnetic exchange $J_H$ between the localized moments. 
Even for quite small $J_H$, a superconducting dome appears between the antiferromagnet and the paramagnetic metals, 
for $n_c \gtrsim 0.8$, see Fig.~\ref{fig:phase-diagram}. We also have indications for a coexistence of magnetism and
superconductivity when the Kondo exchange is small.

Therefore, both the occurrence of a superconducting dome right in the vicinity of the quantum critical point 
separating the magnetic metal from the paramagnetic one, the typical example being 
CePd$_2$Si$_2$,~\cite{mathur1998,grosche2001} and the coexistence of antiferromagnetism and superconductivity, 
observed in CeRhSi$_2$,~\cite{movshovich1996} CeRhIn$_5$,~\cite{yashima2007} and, more recently, 
CeCo(In$_{1-x}$Cd$_x$)$_5$,~\cite{nair2010} are reproduced by an enriched Kondo lattice model.

This work was partially supported by PRIN 2010-11.


\begin{thebibliography}{99}
\bibitem{steglich1979} F. Steglich, J. Aarts, C.D. Bredl, W. Lieke, D. Meschede, W. Franz, and 
   H. Schafer, \prl {\bf 43}, 1892 (1979).
\bibitem{norman2011} M.R. Norman, Science {\bf 332}, 196 (2011). 
\bibitem{scalapino2012} D.J. Scalapino, \rmp {\bf 84}, 1383 (2012). 
\bibitem{allan2013} M.P. Allan, F. Massee, D.K. Morr, J. Van Dyke, A.W. Rost, A.P. Mackenzie, 
   C. Petrovic, and J.C. Davis, Nat. Phys. {\bf 9}, 468 (2013).
\bibitem{mathur1998} N.D. Mathur, F.M. Grosche, S.R. Julian, I.R. Walker, D.M. Freye, 
   R.K.W. Haselwimmer, and G.G. Lonzarich, Nature (London) {\bf 394}, 39 (1998).
\bibitem{grosche2001} F.M. Grosche, I.R. Walker, S.R. Julian, N.D. Mathur, D.M. Freye, 
   M.J. Steiner, and G.G. Lonzarich, J. Phys.: Condens. Matter {\bf 13}, 2845 (2001).
\bibitem{si2010} Q. Si and F. Steglich, Science {\bf 329}, 1161 (2010).
\bibitem{movshovich1996} R. Movshovich, T. Graf, D. Mandrus, J.D. Thompson, J.L. Smith, and 
   Z. Fisk, \prb {\bf 53}, 8241 (1996).
\bibitem{yashima2007} M. Yashima, S. Kawasaki, H. Mukuda, Y. Kitaoka, H. Shishido, R. Settai, 
   and Y. Onuki, \prb {\bf 76}, 020509 (2007).
\bibitem{nair2010} S. Nair, O. Stockert, U. Witte, M. Nicklas, R. Schedler, K. Kiefer, 
   J.D. Thompson, A.D. Bianchi, Z. Fisk, S. Wirth, and F. Steglich, PNAS {\bf 107}, 9537 (2010).
\bibitem{doniach1977} S. Doniach, Physica B \& C {\bf 91}, 231 (1977).
\bibitem{tsunetsugu1997} H. Tsunetsugu, M. Sigrist, and K. Ueda, \rmp {\bf 69}, 809 (1997).
\bibitem{coleman2007} P. Coleman, in {\it Handbook of Magnetism and Advanced Magnetic Material}, 
   Volume 1: Fundamentals and Theory, pag. 95 (eds H. Kronmuller and S. Parkin) (Wiley, 2007).
\bibitem{anderson1987} P.W. Anderson, Science {\bf 235}, 1196 (1987).
\bibitem{xavier2003} J.C. Xavier, \prb {\bf 68}, 134422 (2003).
\bibitem{asadzadeh2013} M.Z. Asadzadeh, F. Becca, and M. Fabrizio, \prb {\bf 87}, 205144 (2013).
\bibitem{bodensiek2013} O. Bodensiek, R. Zitko, M. Vojta, M. Jarrell, and T. Pruschke, \prl {\bf 110},
    146406 (2013).
\bibitem{coleman2010} P. Coleman and A.H. Nevidomskyy, J. of Low Temp. Phys. {\bf 161}, 182 (2010).
\bibitem{nakatsuji2006} S. Nakatsuji, Y. Machida, Y. Maeno, T. Tayama, T. Sakakibara, J. van Duijn,
   L. Balicas, J.N. Millican, R.T. Macaluso, and J.Y. Chan, \prl {\bf 96}, 087204 (2006).
\bibitem{kim2008} M.S. Kim, M.C. Bennett, and M.C. Aronson, \prb {\bf 77} 144425 (2008).
\bibitem{motome2010} Y. Motome, K. Nakamikawa, Y. Yamaji, and M. Udagawa, \prl {\bf 105}, 036403 (2010).
\bibitem{bernhard2011} B.H. Bernhard, B. Coqblin, and C. Lacroix, \prb {\bf 83}, 214427 (2011).
\bibitem{peters2011} R. Peters, N. Kawakami, and T. Pruschke, J. Phys.: Conf. Ser. {\bf 320} 012057 
   (2011).
\bibitem{rau2014} J.G. Rau and H.-Y. Kee, \prb {\bf 89}, 075128 (2014).
\bibitem{wu2014} W. Wu and A.M.S. Tremblay, arXiv:1410.1496.
\bibitem{coleman1989} P. Coleman and N. Andrei, J. Phys.: Condens. Matter {\bf 1}, 4057 (1989).
\bibitem{iglesias1997} J.R. Iglesias, C. Lacroix, and B. Coqblin, \prb {\bf 56}, 11820 (1997);
   B. Coqblin, C. Lacroix, M.A. Gusmao, and J.R. Iglesias, \prb {\bf 67}, 064417 (2003).
\bibitem{kim2003} M.D. Kim, C.K. Kim, and J. Hong, \prb {\bf 68}, 174424 (2003).
\bibitem{xavier2008} J.C. Xavier and E. Dagotto, \prl {\bf 100}, 146403 (2008).
\bibitem{liu2012} Y. Liu, H. Li, G.-M. Zhang, and L. Yu, \prb {\bf 86}, 024526 (2012).
\bibitem{isaev2013} L. Isaev and I. Vekhter, \prl {\bf 110}, 026403 (2013).
\bibitem{sato2001} N.K. Sato, N. Aso, K. Miyake, R. Shiina, P. Thalmeier, G. Varelogiannis, 
   C. Geibel, F. Steglich, P. Fulde, and T. Komatsubara, Nature (London) {\bf 410}, 340 (2001).
\bibitem{ogata2007} H. Watanabe and M. Ogata, \prl {\bf 99}, 136401 (2007).
\bibitem{aliaga2001} H. Aliaga, B. Normand, K. Hallberg, M. Avignon, and B. Alascio, \prb {\bf 64}, 024422 (2001).
\bibitem{garcia2004} D.J. Garcia, K. Hallberg, B. Alascio, and M. Avignon, \prl {\bf 93}, 177204 (2004).
\bibitem{sorella2005} S. Sorella, \prb {\bf 71}, 241103 (2005).
\end{thebibliography}
\end{document}